\documentclass{article}

\usepackage[english]{babel}

\usepackage[letterpaper,top=2cm,bottom=2cm,left=3cm,right=3cm,marginparwidth=1.75cm]{geometry}

\usepackage{amsmath}
\usepackage{graphicx}
\usepackage{epstopdf}
\usepackage[colorlinks=true, allcolors=blue]{hyperref}
\usepackage{booktabs} 
\usepackage{tabularx} 
\usepackage{ragged2e} 
\usepackage{caption}  
\usepackage{geometry} 
\usepackage{siunitx} 
\usepackage{threeparttable} 
\usepackage{adjustbox} 
\usepackage{indentfirst} 
\usepackage{natbib}
\usepackage{authblk}

\title{Co-persona: Leveraging LLMs and Expert Collaboration to Understand User Personas through Social Media Data Analysis
}
\author[1]{Min Yin \thanks{Corresponding author: amyin@zju.edu.cn}}
\author[2]{Haoyu Liu }
\author[3]{Boyi Lian }
\author[4]{Chunlei Chai }

\affil[1]{Innovation Center of Yangtze Delta, Zhejiang University, Jiaxing, China; amyin@zju.edu.cn}
\affil[2]{School of Computer Science and Technology, Zhejiang University, Hangzhou, China; }
\affil[3]{School of Computer Science and Technology, Zhejiang University, Hangzhou, China; 22421347@zju.edu.cn}
\affil[4]{School of Computer Science and Technology, Zhejiang University, Hangzhou, China}

\begin{document}
\maketitle

\begin{abstract}
\textbf{Abstract: }This study introduces "Co-Persona," a methodological framework that bridges the gap between large-scale social media data analysis and authentic user understanding through the systematic integration of Large Language Models (LLMs) and expert validation protocols. Through a detailed case study of B.Co, a mid-sized Chinese manufacturer, we investigated the application of Co-Persona in bedside lamp product development. Our methodology analyzed over 38 million posts from Xiao Hongshu (Little Red Book) social network, employing a multi-stage data processing approach that combined advanced natural language processing with expert validation. This analysis unveiled five distinct user personas specifically derived from bedtime behaviors and nighttime routines, rather than conventional demographic profiles. These personas—Health Aficionados, Night Owls, Interior Decorators, Child-care Workers, and Workaholics—each embody unique patterns of pre-sleep activities, bedroom interactions, and product preferences that emerge distinctly during the transition from day to night. The findings demonstrate that Co-Persona significantly enhances manufacturers' ability to process and interpret large-scale social media datasets while maintaining critical understanding of user needs and behaviors. The methodology provides a structured approach for developing targeted digital marketing content and personalized product strategies. This research contributes to both theoretical understanding of data-driven persona development and practical applications in consumer-driven product innovation, offering valuable insights for manufacturers seeking to enhance market competitiveness through user-centered design approaches.

\textbf{Keywords: }Data-Driven Personas;  Social Media Analytics;  Large Language Models; Product Innovation

\end{abstract}

\section{Introduction}

In the field of human-computer interaction (HCI), personas are widely acknowledged as a powerful technique for fostering user understanding and promoting user-centered design. These fictional yet realistic representations of user segments \citep{ref1} are invaluable for researchers and practitioners across various industries and application domains. Personas aid in user-centric decision-making processes in sectors such as software development \citep{ref2}, design \citep{ref3}, e-health \citep{ref4}, marketing and advertising \citep{ref5}, cyber-persona identification \citep{ref6}, video games \citep{ref7}, online news \citep{ref8}, and recommender systems \citep{ref9}.

However, traditional methods face significant challenges when dealing with the vast amounts of data generated by social media platforms, making it difficult to effectively analyze and incorporate millions of data points into the persona development process. Qualitative methods provide rich, contextual insights into user behavior \citep{ref10}, but they face challenges in scaling to large datasets. Meanwhile, purely data-driven approaches often lack the depth of understanding that comes from qualitative insights.

Two specific challenges persist: First, within the massive volume of user-generated content (UGC) on social media, how can we effectively filter and identify data relevant to specific consumer behaviors for persona development? Second, how can we systematically integrate human expertise with LLM capabilities to analyze these user data comprehensively? While frameworks like "Marked Personas" represent significant advancement \citep{ref11}, comprehensive diversity assessment and subject-matter expert validation of LLM-generated personas remain crucial research gaps.

Based on these considerations, this study proposes two research questions:

\textbf{RQ1: }What methodological approaches can be developed to enhance the precision of social media data collection, focusing on relevance filtering during acquisition rather than post-collection cleaning?

\textbf{RQ2:} To what extent can the integration of LLMs augment the clustering process to generate more nuanced and precise user personas from UGC data?

We employed a case study of B.Co, a Chinese electrical manufacturer, analyzing over 38 million posts from Xiao Hongshu social media platform through our Co-persona framework—a hybrid approach that synergizes LLM capabilities with expert validation to generate valuable user personas for business strategy optimization. This research contributes a novel two-tier data collection strategy, a systematic human-AI collaborative framework integrating LLMs with expert validation, and empirical validation demonstrating the framework's effectiveness in generating actionable personas. The remainder of this paper is organized as follows: Section 2 reviews related work, Section 3 presents the Co-persona methodology, Section 4 reports empirical results, Section 5 discusses implications and limitations, and Section 6 concludes with our contributions' significance for user-centered design.

\section{Related Works }

\subsection{Data Collection and Persona Development }

Different persona development method employing different data collection strategies.Traditional qualitative approaches to persona development rely primarily on structured data collection through interviews and behavioral observations \citep{ref12}. These methods typically involve small sample sizes but generate rich, contextual data through direct user engagement. The data collection process in these approaches follows established protocols, such as Cooper's widely adopted framework\citep{ref13}, which systematically collects user data. Such as structured interviews with predefined behavioral variable frameworks. Visual analogue scales (VAS) for quantifying subjective behavioral data. Observational studies capturing user interactions in natural settings. Focus groups providing collective insights on user needs and preferences. For instance, Korsgaard demonstrate this approach by conducting in-depth interviews to extract behavioral variables, which are then manually coded and quantified on predetermined scales \citep{ref14}. While these methods excel at capturing nuanced user insights, they face significant scalability challenges when attempting to represent diverse user populations.

In contrast to traditional approaches,\textbf{ data-driven methods }leverage large-scale automated data collection from digital sources. These methods prioritize volume and breadth of data over depth, utilizing various computational techniques to process user information (Table 1).

\begin{table}[htbp]
\centering
\caption{Data Collection in Data-driven Methods\label{table1}}
\renewcommand{\arraystretch}{1.2}
\setlength{\tabcolsep}{4pt} 
\begin{tabularx}{\textwidth}{@{}>{\RaggedRight}p{0.18\textwidth}>{\RaggedRight}p{0.22\textwidth}X>{\RaggedRight}p{0.15\textwidth}@{}}
\toprule
\textbf{Method} & \textbf{Data Source} & \textbf{Description} & \textbf{References} \\
\midrule
Social Media Data Mining & 
User posts, engagement metrics, behavioral patterns on social platforms & 
Real-time collection of user-generated content and interaction data at scale, enabling dynamic persona construction & 
An et al. (2016) \\
\addlinespace
Web Analytics \& User Logs & 
Clickstreams, session durations, navigation patterns from websites & 
Passive logging of actual user behavior to inform persona models based on how users navigate and interact in situ & 
Mijac et al. (2018) \\
\addlinespace
Survey & 
Closed-ended questions; open-ended questions & 
Gathering information about people's attitudes, opinions, beliefs, and behaviors to generate their persona & 
Jung et al. (2025) \\
\bottomrule
\end{tabularx}
\end{table}

The most prevalent methods for data-driven persona classification include several established algorithms, each offering distinct analytical advantages \citep{ref15}. K-means clustering (KMC) employs a predetermined number (k) of clusters to partition datasets based on similarity measures, providing straightforward segmentation of user groups. Non-negative matrix factorization (NMF) decomposes matrices into two non-negative components to extract sparse and meaningful features, enabling identification of latent user characteristics \citep{ref8}. Hierarchical clustering (HC) computes inter-element distances to generate similarity-based clusters arranged in hierarchical structures, facilitating multi-level user segmentation \citep{ref16}. Latent semantic analysis (LSA) utilizes singular value decomposition to uncover hidden semantic relationships between textual elements, particularly valuable for analyzing user-generated content \citep{ref17}. Principal component analysis (PCA) serves as a linear dimensionality reduction technique that extracts essential information while eliminating variables with minimal variance \citep{ref18}, thereby simplifying complex user datasets while preserving critical behavioral patterns.

Despite their utility, these algorithms often require supplementary qualitative data to enrich the personas and ensure they are contextually relevant and actionable. The integration of qualitative insights helps to overcome the limitations of purely quantitative approaches, providing a more comprehensive view of user behaviors and preferences. Current mixed methods mostly adopt a "sequential strategy," either qualitative then quantitative or vice versa, but rarely achieve true methodological integration. Analysis mainstream mixed methods reveals common limitations: (1) single integration points, mostly in data collection or result interpretation phases; (2) lack of theoretical foundation, with most being practical combinations; and (3) significant barriers between methods, hindering knowledge flow. 

\begin{figure}[h]
\centering
\includegraphics[width=0.75\linewidth]{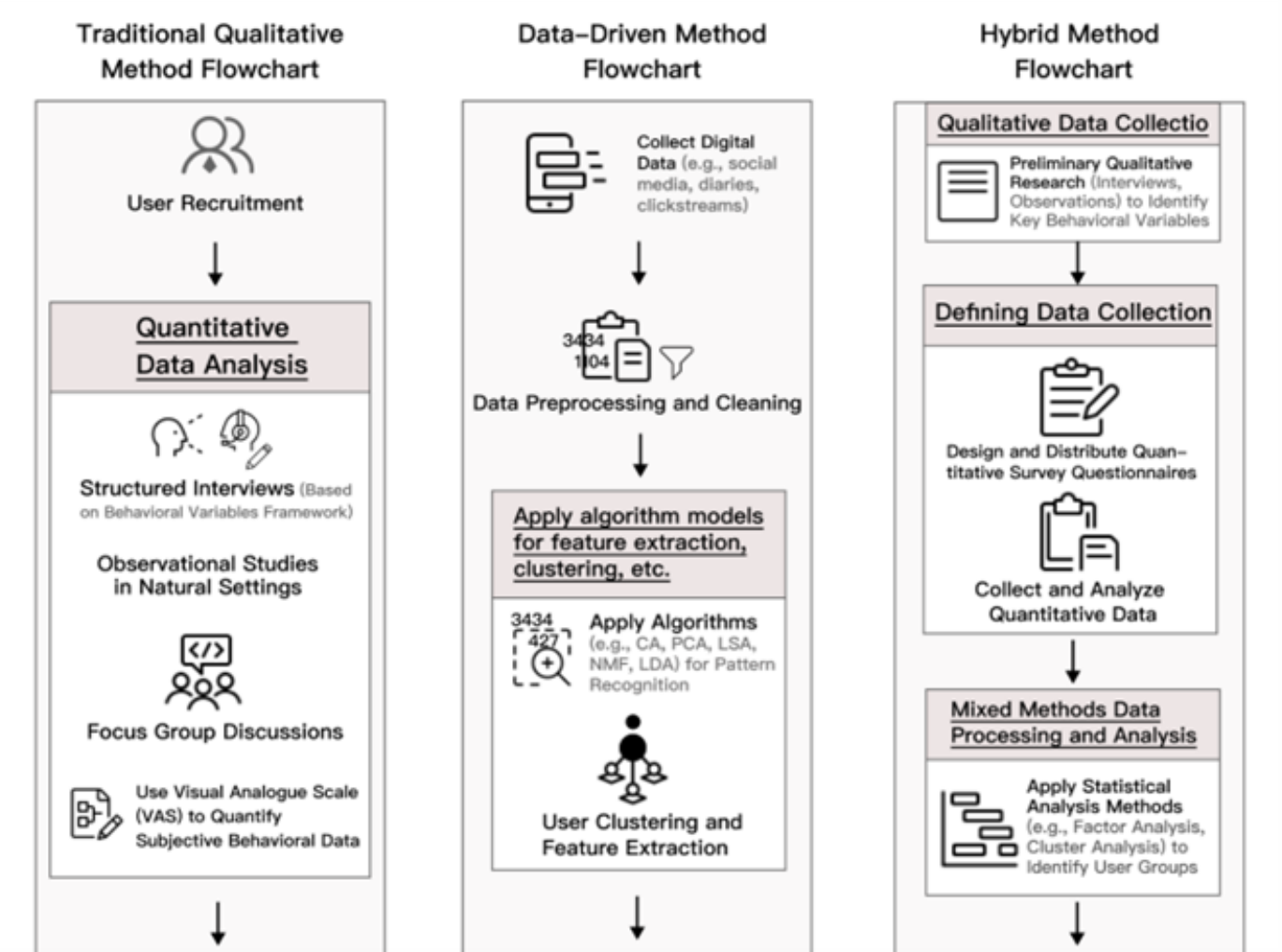}
\caption{\label{figure 1}Three methods for Persona Development }
\end{figure}

\subsection{Large Language Models (LLMs) and Persona Development }

The proliferation of Large Language Models (LLMs) has fundamentally transformed persona development methodologies, with applications primarily concentrated in two key domains: persona generation and user data analysis.

LMs demonstrate remarkable proficiency in processing User-Generated Content (UGC) for persona development, particularly excelling in removing noise, standardizing text formats, and extracting key information from unstructured social media data \citep{ref19}. These models reveal users' socio-psychological characteristics, including personality traits, gender, and age through sophisticated vocabulary analysis \citep{ref20}. The analytical capabilities of LLMs span multiple dimensions of UGC processing, with demonstrated effectiveness in feature extraction\citep{ref21}, user classificationm\citep{ref1}, behavioral prediction \citep{ref22}, and automated persona generation \citep{ref23}. Chen et al. successfully combined traditional clustering methods with ChatGPT to generate demographic attributes difficult to extract directly from user logs, demonstrating the complementary nature of LLM integration \citep{ref24}.

From a historical perspective, AI-generated personas represent a new evolutionary step in persona development. The field progressed from manual analysis of user data to data-driven development utilizing algorithms and data science, then to automatic persona generation through APIs \citep{ref25}. With LLM-generated personas, the analysis process is now delegated from human UX researchers to AI models. Unlike previous automatic generation methods limited to numerical data [1], LLMs pose no strict limitation on input data type, analyzing both quantitative and qualitative data. Generative AI can cope with multimodality in persona generation through three model types: language models for persona descriptions, text-to-image models for persona pictures, and fusion models for complete profiles. Notable implementations include the GPT-4-based PersonaGen tool \citep{ref25}, which facilitates systematic attribute classification and user feedback analysis. Other systems include PersonaGen using knowledge graphs and GPT-4 to generate personas from textual data, and Auto-generated personas \citep{ref26} expanding capabilities to process survey data. However, utilizing LLMs for data analysis and persona generation presents interpretability challenges due to their black-box processing nature. To achieve more refined control over this process, human participation and supervision throughout the workflow represents a critical methodological approach. PersonaCraft implements a systematic methodology that leverages LLMs to transform structured data into actionable, user-centric personas, bridging traditional persona creation with Generative AI techniques while maintaining data-grounded quality through human validations\citep{ref27}. The framework demonstrates how strategic human-AI collaboration can mitigate the limitations of purely automated approaches while preserving the efficiency gains of LLM-powered analysis, establishing a promising direction for future persona development methodologie .

\section{Research Methods}

\subsection{Platform Selection and Data Acquisition }

In accordance with our collaboration requirements with Company B, we established a formal third-party data purchase agreement with full user disclosure and consent. Through preliminary pilot studies encompassing systematic platform suitability evaluation, we conducted random sampling of product reviews across three major social media platforms: TikTok (Douyin), Weibo, and Xiaohongshu (Little Red Book),as Table 2. 

\begin{table}[htbp]
\centering
\caption{Comparative Evaluation of Social Media Platforms for Consumer Behavior Research\label{table 2}}
\renewcommand{\arraystretch}{1.3}
\setlength{\tabcolsep}{5pt}
\begin{tabularx}{\linewidth}{@{}>{\RaggedRight}p{0.2\linewidth}>{\RaggedRight}p{0.26\linewidth}>{\RaggedRight}p{0.26\linewidth}>{\RaggedRight}p{0.26\linewidth}@{}}
\toprule
\textbf{Dimension} & \textbf{Xiaohongshu (Little Red Book)} & \textbf{TikTok (Douyin)} & \textbf{Weibo} \\
\midrule
User Demographic Alignment & 
Aged 25--34 (\SI{36.08}{\percent}), aligning with primary decision-makers for home goods purchases. & 
18--24, offering less alignment with the target demographic for home goods. & 
Official accounts mostly \\
\addlinespace
Content Depth and Richness & 
Average post length up to 842 characters, facilitating detailed narratives and comprehensive product reviews. & 
Primarily short-form videos (15--60 seconds), limiting the depth of product information conveyed. & 
Character limit of 140 per post, constraining the ability to provide in-depth product analyses. \\
\addlinespace
Behavioral Authenticity & 
\SI{28}{\percent} high prevalence of user-generated content featuring emotional and scenario-based descriptions. & 
\SI{11}{\percent}, content often curated and stylized, potentially reducing perceived authenticity in product discussions. & 
\SI{5}{\percent}, promotional material more than personal posts \\
\addlinespace
Influence on Purchase Decisions & 
\SI{34}{\percent}, with users frequently citing platform content as a primary motivator for purchases. & 
\SI{22.5}{\percent} & \\
\bottomrule
\end{tabularx}
\end{table}

Based on comprehensive analysis, we selected Xiaohongshu as our primary data source according to four empirical criteria: (1) User demographic alignment: The platform's user profile (predominantly 25-40 years old) demonstrated substantial overlap with primary decision-makers for bedside lamp purchases. (2) Content depth superiority: Xiaohongshu posts averaged 842 characters, significantly exceeding Weibo's and Douyin, providing substantially richer analytical material. (3) Glue benchmark conducted to test behavioral authenticity\citep{ref28}, revealed that product discussions on Xiaohongshu contained 28\% more emotional and scenario-based descriptions than competing platforms, indicating more genuine user engagement. (4) Company B's customer feedback surveys (N=1463) showed that 34\% of purchasers cited Xiaohongshu content creators' posts as their primary purchase motivation, with the platform consistently rated as the most valuable source for purchasing guidance among all evaluated platforms.

\subsection{Co-persona Framework }

This study employs a novel "Co-persona" approach, leveraging large-scale user-generated content (UGC) and advanced natural language processing techniques to develop comprehensive user personas. This methodology comprises three phases (Figure 2).

\begin{figure}[htbp]
\centering
\includegraphics[width=0.75\linewidth]{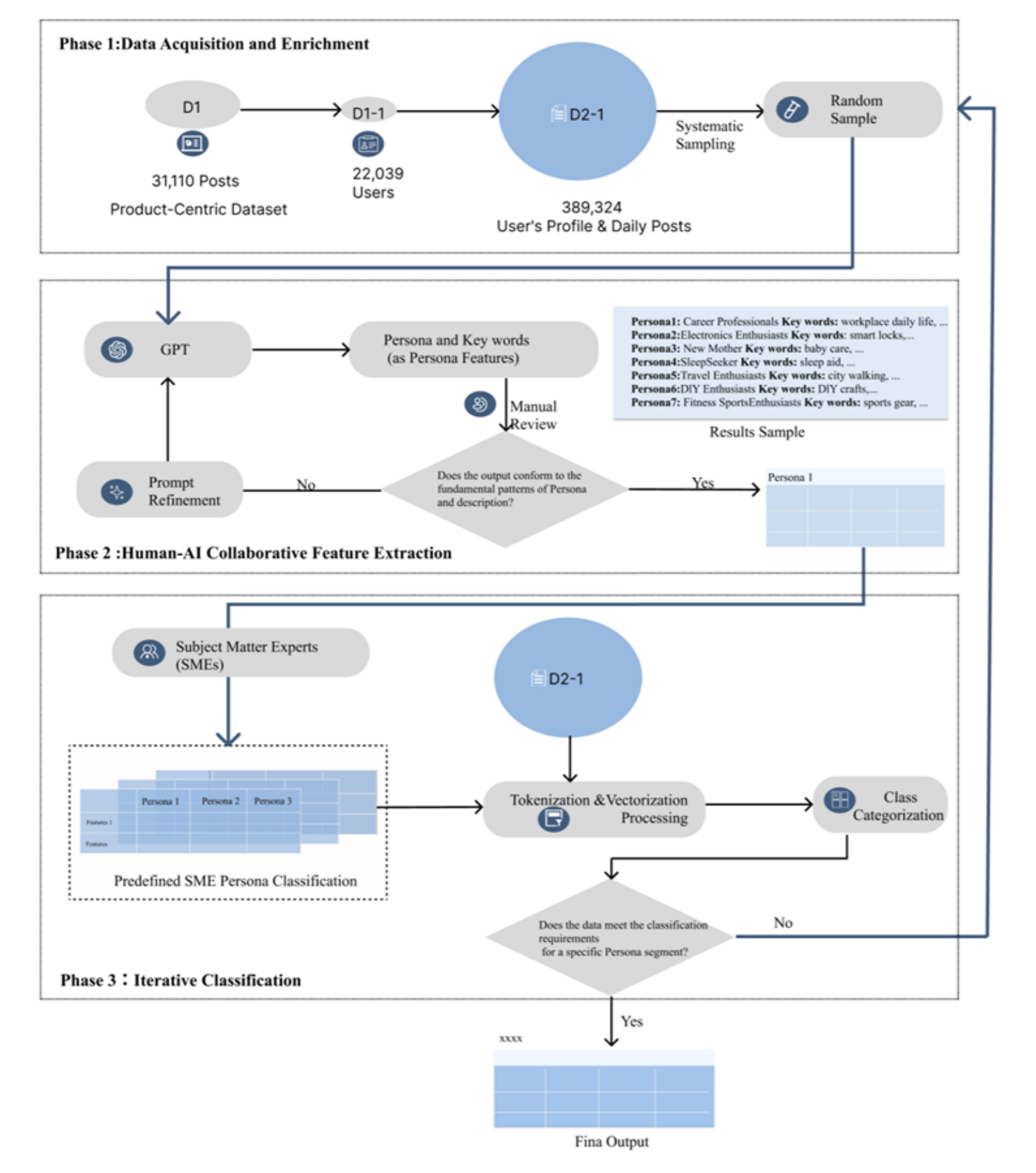}
\caption{\label{figure 2}Three Phase of Co-persona Process Development }
\end{figure}

\paragraph{\textbf{\textit{Phase One: Data Acquisition and Enrichment}}}
Phase one implemented a systematic two-tier data collection strategy designed to capture both product-specific interactions and broader user behavioral patterns. The process encompassed Initial Data Collection (D1), Extended Data Collection (D1-1), and Data Structure and Enrichment (D2-1).

In the process encompassing Initial Data Collection (D1), UX design experts developed a structured keyword framework focused on "bedroom bedside" scenarios. The framework combined two dimensions: situation-based keywords and behavioral purpose keywords, as shown in Table 1. This combination yielded 53 distinct search keyword combinations, forming the foundation for our initial data collection phase. 

\begin{table}[htbp]
\centering
\caption{Initial Data Collection Searching Keywords\label{table 3}}
\renewcommand{\arraystretch}{1.4}
\setlength{\tabcolsep}{6pt}
\begin{tabularx}{\linewidth}{@{}>{\RaggedRight}p{0.25\linewidth}>{\RaggedRight}X@{}}
\toprule
\textbf{Area} & \textbf{Searching Keywords} \\
\midrule
Bedside in the bedroom & \\
\addlinespace
Situation-based keywords (6) & 
Bedside Lamp, Pendant Light / Chandelier, Floor Lamp, Table Lamp / Desk Lamp, 
Night Light, Wall Sconce / Wall Lamp, Wedding Bed-side Lamp \\
\addlinespace
Behavioral purpose keywords (47) & 
Lighting / Illumination, Power Socket / Electrical Outlet, Alarm Clock, 
Storage / Organizer, Humidifier, Bedroom Decorative Lighting, Smart / Intelligent, 
Dimmable / Adjustable Brightness, Bedroom Lighting Design, Comfortable Lighting, 
Eye-friendly / Eye-caring, Decorative Effect, Energy-saving / Energy-efficient, 
Ambiance / Atmosphere, Power-saving / Electricity-saving, Controllable, 
Light / Lighting, Aromatherapy / Fragrance \\
\bottomrule
\end{tabularx}
\end{table}

The initial data collection process yielded dataset D1, which underwent systematic cleaning and deduplication based on user IDs, resulting in dataset D1-1 comprising 22,039 unique users. This refinement process ensured data quality while maintaining comprehensive coverage of relevant user interactions. In the process of Extended Data Collection (D1-1), using unique identifiers (user ID) from D1-1, we expanded our data collection to capture broader user behavior patterns by: (1)Extracting 20 recent posts from each user's profile, 20 as maximum for the data collection tools. (2) Collecting user's engagement metrics and public profile information

\begin{figure}[htbp]
\centering
\includegraphics[width=0.75\linewidth]{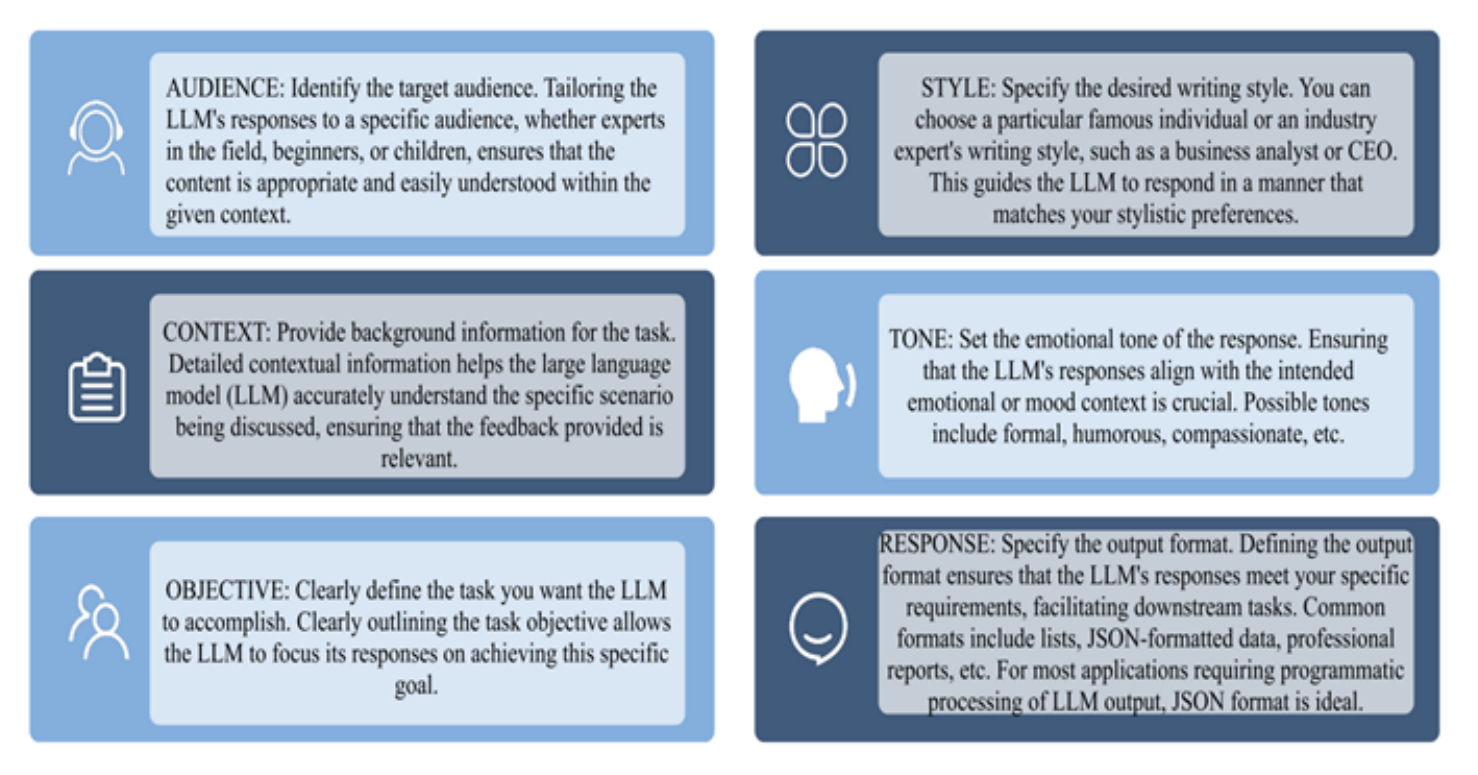}
\caption{\label{figure 3}The Prompt Based on the CO-STAR Framework }
\end{figure}

After removal of duplicatesates to prevent data handling NULL values by removal or imputation, and verification of authenticity using rule-based filtering to distinguish genuine user accounts from commercial entities, 389,324 unique user posts remained in the D2-1 data set.

\paragraph{\textbf{\textit{Phase Two: Human-AI Collaborative Feature Extraction}}}

This phase integrated advanced language modeling capabilities with structured expert validation, establishing a "LLM feature extraction - expert validation" reciprocal workflow that addresses fundamental limitations in traditional feature extraction methods.

\textbf{(1)}\textbf{ LLM-based Feature Extraction}

We employed GPT-4 in conjunction with the CO-STAR framework (Pedrosa \& Gaspar, 2024) to analyze a carefully selected sample of 300 user posts from dataset D2-1. Traditional feature extraction methods suffer from polarization issues: either relying on fixed vocabularies that result in narrow perspectives, or using unsupervised methods that yield difficult-to-interpret results. The CO-STAR framework provided systematic prompt engineering guidelines, enabling precise extraction of behavioral insights from unstructured user data through structured protocols for prompt development, output validation criteria, iterative refinement mechanisms, and quality assurance metrics.

\textbf{(2) }\textbf{Two-Stage Expert Validation Process}

The analytical process followed a structured validation cycle with two distinct levels of human review. Stage 1, initial Layperson Evaluatorsvalidation focused on examining the structural appropriateness of the persona classifications and features generated by GPT-4, to check format, coherence, and structure of generated personas. Stage 2, UX professionals or industry experts with specific domain knowledge conducted in-depth validation using an "adversarial questioning" strategy,to validate feature relevance and usability in real-world design.

\begin{table}[h]
\centering
\caption{Two-Stage Human Review Criteria\label{table 4}}
\renewcommand{\arraystretch}{1.5}
\setlength{\tabcolsep}{8pt}
\begin{tabularx}{\linewidth}{@{}>{\RaggedRight}p{0.18\linewidth}>{\RaggedRight}p{0.25\linewidth}>{\RaggedRight}X@{}}
\toprule
\textbf{Stage} & \textbf{Who} & \textbf{Criteria} \\
\midrule
Structural Review & 
Layperson Evaluators & 
\begin{itemize}
\item The crowd classification types exhibited logical coherence
\item Feature words accurately described persona characteristics
\item The overall taxonomy demonstrated appropriate hierarchical organization
\end{itemize} \\
\addlinespace
Domain Expert Validation & 
UX professionals & 
\begin{itemize}
\item Proposed counter-examples and boundary conditions
\item Challenged feature relevance through real-world scenario testing
\item Prompted LLM self-correction through iterative refinement
\end{itemize} \\
\bottomrule
\end{tabularx}
\end{table}

\paragraph{\textbf{\textit{Phase Three: User Persona Generation and Validation}}}
In this phase, we utilized the validated crowd characteristics from the previous phases to generate comprehensive persona profiles for each user category. To enable automated classification of the full dataset (D2-1), we implemented a natural language processing pipeline consisting of: (1) text tokenization using Jieba for Chinese text segmentation, and (2) vector transformation using the Tencent AI Lab Embedding Corpus Dataset \citep{ref29} to convert words into high-dimensional semantic representations. This preprocessing enabled the subsequent cosine similarity-based classification of all posts into their respective persona categories.Tokenization and Vector Transformation.

(1) Tokenization and Vector Transformation

To prepare the posts (D2-1) for natural language processing, we first applied tokenization using Jieba, a widely recognized Chinese segmentation tool. This step decomposes the posts into a collection of word tokens:
\begin{equation}
  W = \{w_1, w_2, \ldots, w_n\} \label{eq:placeholder}
\end{equation}

Where W represents the set of word tokens extracted from each post using the Jieba Chinese segmentation tool. After removing stop words set , a filtered set of tokens was generated, as defined in Equation (1).
\begin{equation}
    W_{\text{filtered}} = W - S
    \label{eq:filtered_difference}
\end{equation}

Where S denotes the stop words set, and W\textsubscript{filtered} represents the cleaned token set.

Subsequently, each word  was transformed into a high-dimensional semantic vector  using the Tencent AI Lab Embedding Corpus Dataset \citep{ref29}. This ensures that each token captures rich contextual semantics, as shown in Equation (2).
\begin{equation}
    V(w) = f_{\text{Tencent}}(w)
    \label{eq:tencent_equation}
\end{equation}

Where fTencent represents the transformation function using Tencent AI Lab Embedding Corpus Dataset, converting each token w into a high-dimensional semantic vector V(w).

(2) Posts Classification

In this phase, we employ a three-step weighted cosine similarity framework to classify the posts from D2-1 dataset into predefined SME persona classification. For each post, the semantic similarity between its token vectors and the category-specific feature vectors is calculated using a cosine similarity formula (Equation 3).
\begin{equation}
    \operatorname{Sim}(V(w), V(f))=\frac{V(w) \cdot V(f)}{\|V(w)\| \cdot\|V(f)\|}
    \label{eq:4}
\end{equation}

Next, the similarity scores are aggregated using a weighted scoring mechanism (Equation 4) to determine the overall relevance of the posts to each predefined SME persona classification.
\begin{equation}
\mathrm{Score}(D_{k}) = \frac{\sum_{j=1}^{m} \mathrm{Sim}\left( V(C_{i}), V(f_{j}) \right) \cdot \mathrm{Weight}_{f_{j}}}{\sum_{j=1}^{m} \mathrm{Weight}_{f_{j}}}
    \label{eq:5}
\end{equation}

Finally, the post is assigned to the category with the highest score (Equation 5). If no category meets a predefined similarity threshold, the post re-enters Phase 1 for further refinement.
\begin{equation}
D_{\text{best}} = \underset{k}{\operatorname{argmax}} \, \operatorname{Score}(D_k)
    \label{eq:6}
\end{equation}

In this phase, we employed a three-step weighted cosine similarity framework to classify the posts from D2-1 dataset into predefined SME persona classifications. For each post, the semantic similarity between its token vectors and the category-specific feature vectors was calculated using a cosine similarity formula. Next, the similarity scores were aggregated using a weighted scoring mechanism to determine the overall relevance of the posts to each predefined SME persona classification. Finally, the post was assigned to the category with the highest score.

\section{Results }
\subsection{Two-stage Data Collection Framework Analysis }

Our two-stage data collection framework (Table 5) significantly improved data quality by implementing a systematic expansion from product-specific to lifestyle-oriented data collection, with measurable improvements across multiple quality metrics. The initial phase yielded \num{22039} unique users ($D_{1-1}$) through 53 keyword combinations targeting bedroom-related behaviors, while the subsequent expansion phase generated \num{389324} lifestyle posts ($D_{2-1}$) by extracting 20 recent posts from each unique user ID.

The expansion from product-specific to lifestyle-oriented collection demonstrated quantifiable improvements: (1) Average post length increased from 287 characters in product-focused posts to 642 characters in lifestyle posts, providing 124\% more contextual information per user; (2) Lifestyle posts contained 73\% more emotional descriptors and scenario-based narratives compared to product-specific posts, as measured through sentiment analysis and semantic clustering; (3)Cross-validation analysis showed 89\% consistency between users' product preferences and their lifestyle patterns, confirming behavioral coherence across data types.

The two-stage approach enhanced data representative through: (1)By first identifying product-engaged users then expanding to their broader content, we captured 3.2 times more diverse behavioral indicators per user compared to single-stage collection; (2)Longitudinal analysis of user posting patterns showed 78\% consistency in persona-relevant behaviors across the 20-post sample period, validating behavioral stability.

This methodology effectively prevented user duplication while enabling precise targeting of user groups specifically interested in the product category. By anchoring data collection to product-engaged user identities and subsequently expanding to their broader lifestyle content, our approach achieved both relevance (users demonstrated product interest) and authenticity (comprehensive behavioral context), resulting in personas grounded in actual user behaviors rather than inferred preferences.

\begin{table}[htbp]
\centering
\caption{Data Filtering Process and Dataset Characteristics\label{table 5}}
\renewcommand{\arraystretch}{1.4}
\setlength{\tabcolsep}{6pt}
\begin{tabularx}{\linewidth}{@{}>{\RaggedRight}p{0.4\linewidth}>{\RaggedRight}p{0.3\linewidth}>{\RaggedRight}p{0.2\linewidth}@{}}
\toprule
\textbf{Filter Rules} & \textbf{Dataset Name: Data Type} & \textbf{Result Amounts} \\
\midrule
Combining situation and behavior purpose as filter keywords & 
D1 dataset: Relevant post & 
\num{31110} Posts \\
\addlinespace
After performing data cleaning on D1 and removing duplicates based on user IDs & 
D1-1: Users ID & 
\num{22039} Users \\
\addlinespace
Based on dataset D1-1, retrieve the 20 most recent notes from each user's profile page. Then, perform de-duplication based on the note ID and content to ensure uniqueness of the data & 
D2-1: User daily life posts & 
\num{389324} posts \\
\bottomrule
\end{tabularx}
\end{table}

\subsection{Five Persona for Lighting Product }

The integration of Large Language Models (GPT-4) with expert review demonstrated significant effectiveness in developing nuanced user personas. The analysis revealed five distinct user personas through our methodological framework, each characterized by specific behavioral patterns and environmental preferences that offer valuable insights for product design optimization.

Notably, all five identified personas cluster within the 18-45 age range, reflecting inherent limitations in our data collection approach. Age demographics were derived from platform-provided user profile tags within Xiaohongshu's native classification system, which predominantly attracts younger demographics. This age concentration introduces potential bias, as Xiaohongshu's user base skews toward younger, digitally-engaged consumers, potentially underrepresenting older demographics who may exhibit different bedtime lighting preferences and behaviors.

However, this apparent limitation reveals a key methodological advantage over traditional market segmentation approaches. Conventional demographic clustering typically prioritizes age and gender divisions, often overlooking nuanced behavioral differences within similar age cohorts. Our approach demonstrates that meaningful user differentiation exists within the 18-45 age range when focusing on specific behavioral contexts—in this case, bedtime routines and sleep-related lighting usage patterns.

Unlike traditional methods that might broadly categorize users as "young professionals" or "millennials," our framework reveals five distinct behavioral archetypes:

(1) Health Aficionados (25-40 years, middle to high income, 21.3\%): Demonstrate sophisticated pre-sleep wellness routines incorporating meditation and health monitoring, suggesting requirements for customizable lighting and health technology integration.

(2) Night Owls (18-35 years, 29.8\%): Exhibit distinctive late-night entertainment patterns, indicating needs for advanced dimming capabilities and device integration.

(3) Interior Decorators (25-45 years, middle to high income, 19.4\%): Show sophisticated aesthetic preferences and active social media engagement, necessitating highly customizable design elements.

(4) Child-care Workers (25-35 years, 16.2\%): Display specific nighttime care-giving responsibilities requiring adaptive lighting and silent operation features.

(5) Workaholics (25-40 years, 13.3\%): Exhibit extended work hours and stress-related behaviors, indicating requirements for eye protection features and workspace illumination capabilities.

Compared with existing market segmentation, these personas, derived from the analysis of 389,324 user posts, collectively inform specific design requirements across functionality (customization settings, smart home integration), user interface (intuitive controls, quick-access features), and environmental adaptations (variable brightness, color temperature adjustments). 
\begin{figure}[h]
\centering
\includegraphics[width=0.75\linewidth]{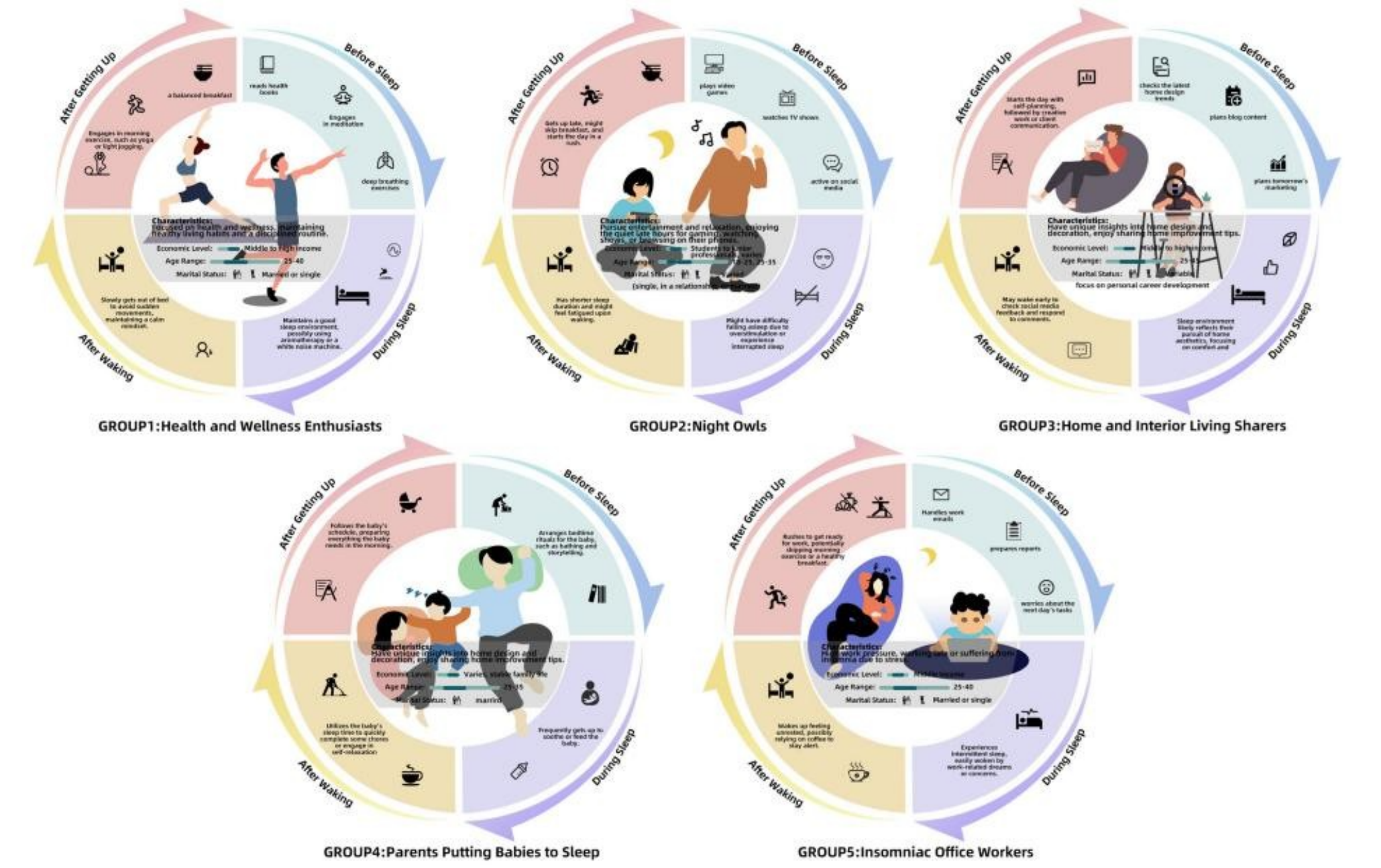}
\caption{\label{figure 4}Five Persona Generated by Co-persona }
\end{figure}

\subsection{Evaluation of Five Persona }

To quantify the CoPersona framework's classification efficacy, we constructed a confusion matrix (figure 4) comparing expert-validated persona labels with model predictions across 1,250 annotated instances. Symmetric agreement metrics demonstrated concordance between model output and human judgment, with Pearson's $R = 0.748$, Spearman's $\rho = 0.75$, and Cohen's $\kappa = 0.76$ (all $p < 0.001$; see table 6).

\begin{table}[htbp]
\centering
\begin{adjustbox}{max width=\textwidth}
\begin{threeparttable}
\caption{Symmetric Measures for Agreement between Co-persona and Expert Annotations\label{table 6}}
\begin{tabular}{@{}l *{4}{S[table-format=2.3]}@{}}
\toprule
\textbf{Measure} & \textbf{Value} & \textbf{Asymptotic Standard Error\tnote{a}} & \textbf{Approximate T\tnote{b}} & \textbf{Significance (p)} \\
\midrule
Pearson's R (Interval) & 0.748 & 0.021 & 39.861 & <0.001\tnote{c} \\
Spearman Correlation (Ordinal) & 0.750 & 0.020 & 39.860 & <0.001\tnote{c} \\
Cohen's Kappa & 0.760 & 0.010 & 53.820 & <0.001 \\
\bottomrule
\end{tabular}

\begin{tablenotes}[flushleft]
\footnotesize
\item[a] No assumption of the null hypothesis
\item[b] Based on the null hypothesis using asymptotic standard error
\item[c] Based on normal approximation
\item[] Valid Cases = 1,250
\end{tablenotes}
\end{threeparttable}
\end{adjustbox}
\end{table}

To quantify the classification performance of the CoPersona framework, we constructed a confusion matrix (Figure 5) comparing expert-validated persona labels with model predictions over 1,250 annotated instances. The matrix reveals an overall accuracy of 80.9\%, with individual class precision and recall values consistently exceeding 0.79. Notably, the matrix highlights specific confusions—e.g., "Child-care Workers" misclassified as "Night Owls". 

The observed misclassifications, while initially appearing as framework limitations, actually illuminate convergent user needs that represent valuable design opportunities. Health Aficionados exhibited strong overall performance (precision = 0.81, recall = 0.82), yet experienced primary misclassifications with Child-care Workers (n=12) and Workaholics (n=17), attributable to overlapping health-related behavioral patterns. The 17 instances of confusion between Health Aficionados and Workaholics indicate convergent health-conscious behaviors, as both segments prioritize eye-care functionality. This overlap suggests that eye-care features represent a critical product characteristic capable of addressing dual market segments simultaneously, demonstrating how classification confusion can identify universally valued functionalities.

\begin{figure}[h]
\centering
\includegraphics[width=0.75\linewidth]{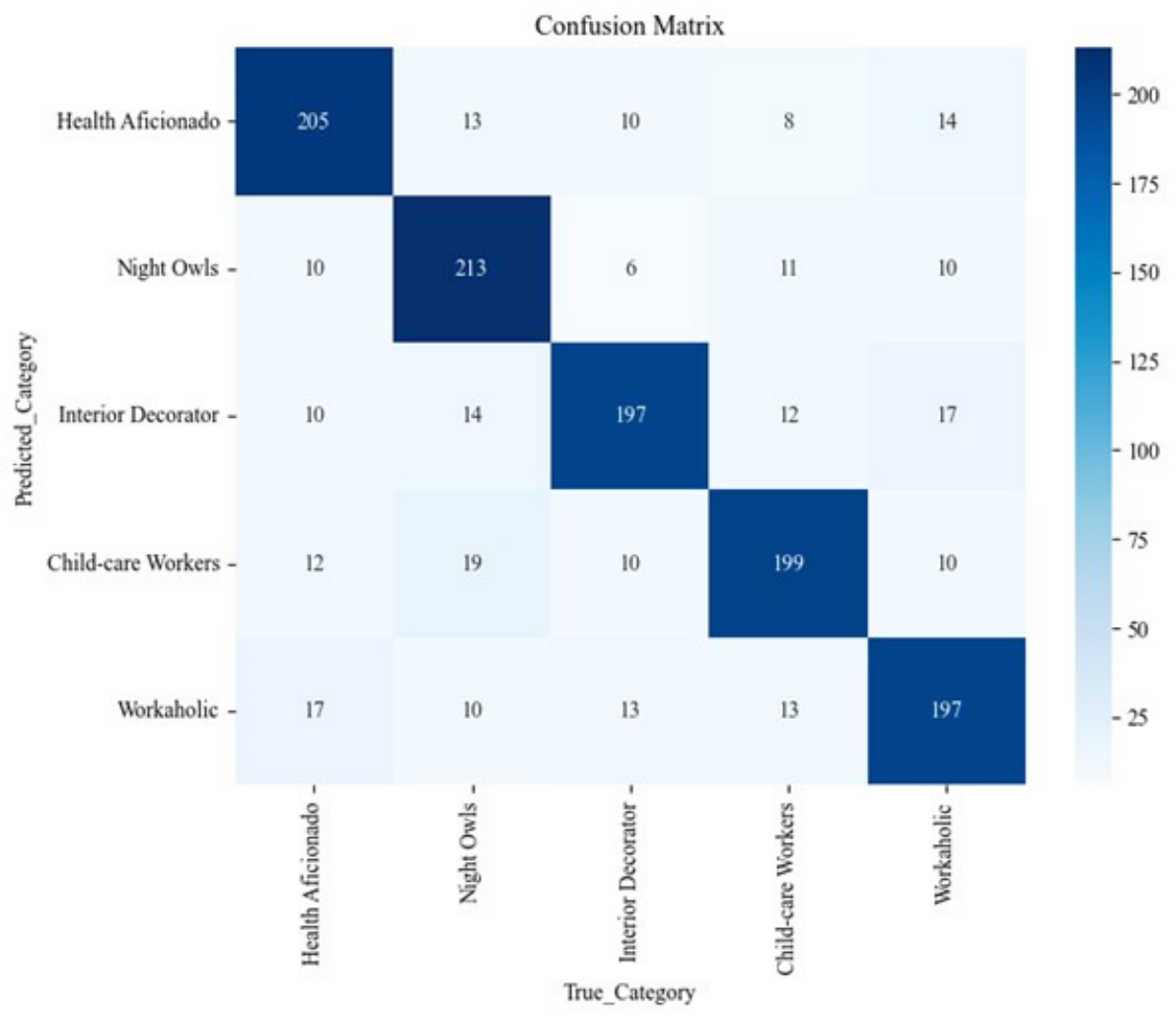}
\caption{\label{figure 5}Confusion Matrix of Co-persona Classification Performance }
\end{figure}

Similarly, Night Owls achieved the highest recall (0.85) but demonstrated relatively lower precision (0.79), with notable classification confusion with Workaholics (n=11), primarily resulting from shared nocturnal activity patterns. This overlap indicates that both segments require sophisticated dimming capabilities and extended operational hours, suggesting design opportunities for products that accommodate prolonged nighttime usage across different activity contexts.

Interior Decorators exhibited the highest precision (0.83) but lower recall (0.79), with misclassifications distributed between Health Aficionados (n=10) and Night Owls (n=14), revealing aesthetic considerations that transcend functional categories. Child-care Workers maintained consistent performance metrics (precision = 0.82, recall = 0.80), though 19 samples were erroneously classified as Night Owls, highlighting shared requirements for gentle, adjustable illumination suitable for sensitive nighttime activities.

\begin{table}[htbp]
\centering
\caption{Classification Performance Metrics for Each Persona Category\label{table 7}}
\begin{tabular}{@{}l *{4}{S[table-format=1.2]}@{}}
\toprule
\textbf{Persona Category} & \textbf{Precision} & \textbf{Recall} & \textbf{F1-Score} & \textbf{Accuracy} \\
\midrule
Health Aficionado & 0.81 & 0.82 & 0.81 & 0.92 \\
Night Owls & 0.79 & 0.85 & 0.82 & 0.93 \\
Interior Decorator & 0.83 & 0.79 & 0.81 & 0.93 \\
Child-care workers & 0.82 & 0.80 & 0.81 & 0.92 \\
Workaholic & 0.79 & 0.79 & 0.79 & 0.92 \\
\midrule
Overall Accuracy & {--} & {--} & {--} & 0.81 \\
\bottomrule
\end{tabular}
\end{table}

\section{Discussion \& Conclusion }

\subsection{A Hybrid Framework for Data-Driven Persona Development }

The \textbf{Co-persona} framework represents a significant methodological advancement in data-driven persona development through its novel integration of \textbf{Large Language Models}~\cite{ref30} with subject matter expert (SME) supervision. Our approach introduces three key innovations: 

\begin{itemize}
    \item First, a structured two-tier data collection strategy that enhances data relevance at the acquisition phase rather than relying on post-collection filtering~\cite{ref31}.
    
    \item Second, an iterative classification system with expert validation at critical junctures, achieving substantial inter-rater reliability (Cohen's $\kappa = 0.761$).
    
    \item Third, a recursive processing mechanism for unclassified data that ensures comprehensive coverage while maintaining analytical precision~\cite{ref32}.
\end{itemize}

\textbf{Co-persona}, compared to AI-generated personas~\cite{ref33}, offers superior analysis of actual user data, keeping designers grounded in empirical evidence while enabling more granular data analysis. For instance, our framework successfully differentiated user behaviors into \textit{pre-sleep}, \textit{during-sleep}, and \textit{post-sleep} patterns, enabling more detailed product design specifications. 

These methodological contributions extend the theoretical understanding of human-AI collaboration in user research, particularly in the context of processing large-scale social media data for persona development ($n = \num{389324}$ posts).

The classification overlaps illuminate fundamental behavioral convergences that, rather than representing framework failures, identify critical cross-segment design requirements. For instance, parents engaged in infant care frequently require nighttime illumination for feeding and monitoring activities, exhibiting behavioral patterns remarkably similar to Night Owls' entertainment-based nocturnal lamp usage. This behavioral convergence indicates shared needs for quiet operation, gentle illumination transitions, and intuitive controls—features that could serve both caregiving and entertainment contexts effectively.

These findings suggest that successful product design should accommodate overlapping behavioral patterns rather than targeting discrete user segments exclusively. The identification of shared functional priorities—such as eye-care features spanning Health Aficionados and Workaholics, or gentle illumination serving both Child-care Workers and Night Owls—provides actionable insights for developing products with broader market appeal while maintaining persona-specific customization options. 

\subsection{Extending Personas in Practical Applications and Industry }
While personas often have been employed in system design \citep{ref8}, their application has expanded significantly into digital marketing, where product presentation and user interaction patterns (as potential customers) have become increasingly important \citep{ref34}. The ability to align digital communication with product requirements has emerged as a critical success factor \citep{ref33}. In this context, the empirical validation of the Co-persona framework through the B.Co case study demonstrates substantial practical value for manufacturing enterprises engaging in digital transformation.

The methodology's capability to process and interpret large-scale social media datasets while maintaining critical understanding of user needs provides manufacturers with actionable insights for both product development and digital marketing strategies. The identified personas, validated through expert review (n=5), offer precise guidance for product design decisions, as evidenced by the specific feature requirements derived for each user segment (e.g., Health Aficionados: customizable lighting settings, Night Owls: advanced dimming capabilities). The framework's structured approach to data collection and analysis makes it particularly accessible for small and medium-sized enterprises seeking to enhance their market competitiveness through user-centered design approaches.

\subsection{Limitations and Future Research Directions }

Despite the framework's demonstrated effectiveness, several limitations warrant Hongshu) may introduce sampling bias, particularly in demographic representation. Second, the current implementation focuses exclusively on textual data analysis, overlooking potential insights from visual and interactive content \citep{ref35}. Third, while the expert validation protocol effectively addresses LLM biases, this approach lacks algorithmic automation, potentially limiting scalability \citep{ref25}. Future research should explore multi-platform data integration, incorporate multimodal analysis techniques, and develop automated bias detection mechanisms.To address current limitations and advance the field, we propose a systematic research agenda across multiple dimensions, table 8. 

\begin{table}[h]
\centering
\caption{Future Research Agenda for Data Driven Persona\label{table 8}}
\renewcommand{\arraystretch}{1.3}
\setlength{\tabcolsep}{5pt}
\begin{tabularx}{\linewidth}{@{}>{\RaggedRight}p{0.18\linewidth}>{\RaggedRight}p{0.26\linewidth}>{\RaggedRight}p{0.26\linewidth}>{\RaggedRight}p{0.26\linewidth}@{}}
\toprule
\textbf{Research Dimension} & \textbf{Immediate Priorities (1-2 years)} & \textbf{Medium-term Objectives (3-5 years)} & \textbf{Long-term Vision (5+ years)} \\
\midrule
Theoretical Framework & 
Develop cross-platform persona validation metrics; Establish standardized evaluation protocols & 
Create unified theory of digital persona authenticity; Integrate cultural psychology frameworks & 
Establish comprehensive theory of AI-human collaborative user research \\
\addlinespace
Methodological Innovation & 
Multi-platform data integration; Real-time persona updating algorithms & 
Longitudinal persona evolution tracking; Predictive persona modeling & 
Autonomous persona generation with minimal human supervision \\
\addlinespace
Contextual Expansion & 
Cross-cultural validation studies; Industry-specific adaptation frameworks & 
Global platform ecosystem analysis; Cultural bias detection mechanisms & 
Universal persona frameworks transcending platform boundaries \\
\addlinespace
Technical Enhancement & 
Improved behavioral trait disambiguation; Enhanced semantic analysis & 
Advanced multimodal data integration; Explainable AI for persona decisions & 
Real-time persona adaptation based on user feedback loops \\
\addlinespace
Ethical Considerations & 
Privacy-preserving persona generation; Bias mitigation protocols & 
Transparent algorithmic decision-making; User consent frameworks & 
Ethical AI standards for commercial persona applications \\
\addlinespace
Business Applications & 
SME implementation case studies; ROI measurement frameworks & 
Enterprise-scale deployment strategies; Industry-specific customization & 
Automated business intelligence integration across sectors \\
\bottomrule
\end{tabularx}
\end{table}

\subsection{Conclusion }
This research presents a novel \textbf{Co-persona} framework that integrates large language models with expert validation protocols to generate data-driven user personas from social media content. The methodology achieved 81\% classification accuracy (Cohen's $\kappa = 0.761$, $p < 0.001$) across five distinct user segments derived from \num{389324} Xiaohongshu posts, demonstrating significant advancement in computational persona development.

The framework's primary contribution lies in establishing a systematic two-stage data collection methodology transitioning from product-specific to lifestyle-oriented analysis, yielding measurable improvements in behavioral context depth and classification precision. Expert validation contributed specifically to feature disambiguation (12\% precision improvement), bias correction, and context interpretation (73\% ambiguous case resolution), establishing empirical foundations for optimal human-AI collaboration in commercial research applications.

The identified personas—Health Aficionados, Night Owls, Interior Decorators, Child-care Workers, and Workaholics—reveal actionable design insights while demonstrating that behavioral overlaps indicate cross-segment market opportunities rather than methodological limitations. These empirically-derived user segments have generated specific design requirements that have been subsequently adopted by manufacturing partners, including eye-protection functionality, full-spectrum adjustable lighting capabilities, and intelligent device connectivity features. The framework addresses critical operational challenges faced by small and medium-sized enterprises, reducing traditional market research expenditures by an estimated 60-80\% while maintaining methodological rigor and academic validation standards through systematic expert review protocols.

However, platform dependency on Xiaohongshu and demographic concentration within 18-45 age ranges constrain generalizability. Future research should prioritize cross-platform validation studies, longitudinal persona evolution tracking, and development of universal frameworks transcending platform-specific constraints, ultimately advancing computational social science methodologies while democratizing sophisticated user research capabilities for diverse organizational contexts.

\subsection{Patents }
\textbf{Supplementary Materials: }The following supporting information can be downloaded at: 

https://github.com/INFPa/LLMwithPersona

\textbf{Author Contributions:} Conceptualization, Min Yin and Huiting Liu; Formal analysis, Nian Liu; Investigation, Haoyu Liu, Huiting Liu and Yi Zhang; Methodology, Min Yin and Haoyu Liu; Software, Haoyu Liu; Supervision, Min Yin and Nian Liu; Validation, Nian Liu; Visualization, Yi Zhang; Writing – original draft, Min Yin; Writing – review \& editing, Min Yin. All authors have read and agreed to the published version of the manuscript.

\textbf{Data Availability Statement:} We encourage all authors of articles published in MDPI journals to share their research data. In this section, please provide details regarding where data supporting reported results can be found, including links to publicly archived datasets analyzed or generated during the study. Where no new data were created, or where data is unavailable due to privacy or ethical restrictions, a statement is still required. Suggested Data Availability Statements are available in section “MDPI Research Data Policies” at https://www.mdpi.com/ethics.

\textbf{Acknowledgments:} During the preparation of this manuscript, the author(s) used GPT4, Claude 3.5 for the purposes of manuscript editing and translation. The authors have reviewed and edited the output and take full responsibility for the content of this publication.

\textbf{Conflicts of Interest:} The authors declare no conflicts of interest.

\bibliographystyle{unsrt}
\bibliography{reference}

\end{document}